\documentclass[doublecol]{epl2}
\usepackage{amssymb}
\usepackage{amsmath}

\shorttitle{}
\shortauthor{F. Author \etal}
\institute{
  \inst{1} Jerusalem College of Engineering, Ramat Beit HaKerem, \\ POB 3566,
Jerusalem, 91035, Israel\\
  \inst{2} Department of Physical Electronics, School of Electrical Engineering,
Faculty of Engineering, Tel Aviv University, Tel Aviv 69978, Israel
}  \abstract{We introduce a system based on
dual-core nonlinear waveguides with the balanced gain and loss
acting separately in the cores. The system features a
"supersymmetry" when the gain and loss are equal to the inter-core
coupling. This system admits a variety of exact solutions (we focus
on solitons), which are subject to a specific subexponential
instability. We demonstrate that the application of a "management",
in the form of periodic simultaneous switch of the sign of the gain,
loss, and inter-coupling, effectively stabilizes solitons, without
destroying the supersymmetry. The management turns the solitons into
attractors, for which an attraction basin is identified. The initial
amplitude asymmetry and phase mismatch between the components
transforms the solitons into quasi-stable breathers.}

\begin{document}

\title{Stabilization of solitons in PT models with supersymmetry by periodic
management}
\author{R. Driben\inst{1,2} \and B. A. Malomed\inst{2}}
\maketitle

\textit{Introduction}. Nonlinear dissipative systems subject to the
parity-time ($\mathcal{PT}$) symmetry feature a unique combination of
conservative and dissipative properties, the most fundamental one being the
real spectrum of eigenvalues generated by non-Hermitian Hamiltonians \cite%
{Bender}. In the simplest case, such a Hamiltonian contains complex
potentials $V(x)$ which obeys the conjugation constraint, $V(x)=V^{\ast
}(-x) $. Recently, it was predicted \cite{Muga, DC1, Berry, DC2, Nimrod,
Longhi, Longhi2, Avadh} and demonstrated in experiments \cite%
{experiment1,Moti} that $\mathcal{PT}$-symmetric systems can be implemented
in optics, making use of symmetric landscapes of the refractive index,
combined with antisymmetrically placed linear amplifiers and absorbers, see
also recent review \cite{DC-review}. The theoretical analysis of $\mathcal{PT%
}$ settings that may be implemented in optics was extended by the
incorporation of nonlinearities \cite{nonlin2, nonlin3, LiKevr, Dmitriev}
(in particular, of cubic gain and loss that are subject to the $\mathcal{PT}$%
-invariance condition in its nonlinear form \cite{Iberia,we}). Solitons were
predicted in some of these nonlinear systems \cite{soliton}.

A setting which may give rise to a $\mathcal{PT}$-symmetric system is a
Kerr-nonlinear dual-core waveguide, with the gain and loss separately
applied to the two cores, which are linearly coupled by the tunneling of
electromagnetic waves across the separating barrier. This setting was
proposed in Ref. \cite{Herb} as a medium capable to support stable
dissipative solitons. Further analysis has corroborated the existence of
one-dimensional solitons in optical \cite{Javid, HS, Pavel, Dima, Chaos} and
plasmonic \cite{Dima} versions of the system. Moreover, stable fundamental
solitons \cite{2D} and vortices \cite{vort,with-Pavel} were predicted in
similarly designed dual laser cavities with saturable nonlinearity. The
advantage of this scheme in comparison with the ordinary (passive) dual-core
systems is that it may be used as the basic element of various lasing setups
\cite{Pavel}-\cite{with-Pavel}. In particular, the $PT$\ symmetry can be
realized in this dual-core system by balancing the gain and loss in the
active and passive linearly-coupled cores. This implies that both cores can
be made \ of a doped lossy material, while the dopant atoms, if pumped
externally, may add a linear gain to the system \cite{China}. Then, if the
external pump is applied to a single core, it may supply the gain
overcompensating the intrinsic loss of this core, so as the net gain
provides the $PT$ \ balance with the other core, that remains lossy. The $PT$%
\ -invariant variant of the active system is especially interesting, as it
corresponds to the lowest gain level which is capable to stabilize the
system.

The dual-core system makes it also possible to define the regime of \emph{%
supersymmetry}, with the strength of the linear coupling between the cores
equal to the common values of the gain and loss coefficients. It is easy to
find a broad class of exact solutions for the dual-core system subject to
the supersymmetry constraint, including solitons. A challenging issue is the
stability of the solitons, as the supersymmetric system is not integrable.
In this work, we find a specific subexponential (resonant) instability to
which the supersymmetric solitons are subject. Then, we demonstrate that the
solitons may be stabilized in the supersymmetric system by means of the
management technique \cite{Sydney,Radik,book}, which periodically switches
the gain and loss between the cores, simultaneously reversing the sign of
the linear coupling. The former element of the management scheme may be
provided by the periodic switch of the core to which the external pump is
applied, while the latter one may actually be implemented by periodically
shifting the phase of one of the field components by $\pi $, i.e.,
periodically passing the wave propagating in one core through $\pi $%
-shifting slabs. The management scheme does not break the supersymmetry,
making the solitons stable in a broad parameter region, which is identified
by means of systematic simulations. The stabilization provided by this
technique complies with the general fact that the management helps stabilize
solitons in a variety of waveguiding systems \cite{Radik,book}.

\textit{The model}. The propagation of optical or plasmon waves in the
dual-core system obeys the linearly coupled equations for the slow evolution
of wave amplitudes $u(Z,T)$ and $v(Z,T)$ in the amplified and damped cores
\cite{Herb}-\cite{Dima}:%
\begin{eqnarray}
iu_{Z}+(1/2)u_{TT}+|u|^{2}u-i\gamma u+\kappa v &=&0,  \notag \\
iv_{Z}+(1/2)v_{TT}+|v|^{2}v+i\Gamma v+\kappa u &=&0.  \label{uv}
\end{eqnarray}%
Here $Z$ is the propagation distance, $T$ is the reduced time or transverse
coordinate in the temporal or spatial system, respectively, $\gamma $ and $%
\Gamma $ are coefficients of the linear gain and loss acting in the two
cores, and $\kappa $ is the strength of the linear coupling between them.
The coefficients in front of the group-velocity dispersion (or diffraction)
and Kerr terms in Eqs. (\ref{uv}) are scaled to be $1$. The $\mathcal{PT}$%
-symmetry takes place for $\Gamma =\gamma $, while the supersymmetry
condition is $\Gamma =\gamma =\kappa $, which casts Eqs. (\ref{uv}) into the
following form:

\begin{eqnarray}
iu_{Z}+(1/2)u_{TT}+|u|^{2}u-i\kappa u+\kappa v &=&0,  \notag \\
iv_{Z}+(1/2)v_{TT}+|v|^{2}v+i\kappa v+\kappa u &=&0.  \label{uv1}
\end{eqnarray}%
Obviously, substitution $v\left( Z,T\right) =iu\left( Z,T\right) $
transforms both equation (\ref{uv1}) into the standard NLS (nonlinear Schr%
\"{o}dinger) equation, $iu_{Z}+(1/2)u_{TT}+|u|^{2}u=0$, hence any solution
to the latter equation generates an exact solution of the supersymmetric
system, the most important one being the fundamental soliton with amplitude $%
\eta $,%
\begin{equation}
u_{\mathrm{sol}}=\eta ~\exp \left( i\eta ^{2}{Z}/4\right) \mathrm{sech}%
\left( \eta T\right) ,~v_{\mathrm{sol}}=iu_{\mathrm{sol}}\text{.}
\label{sol}
\end{equation}

\begin{figure}[tbp]
\centerline{\includegraphics[width=7.5cm]{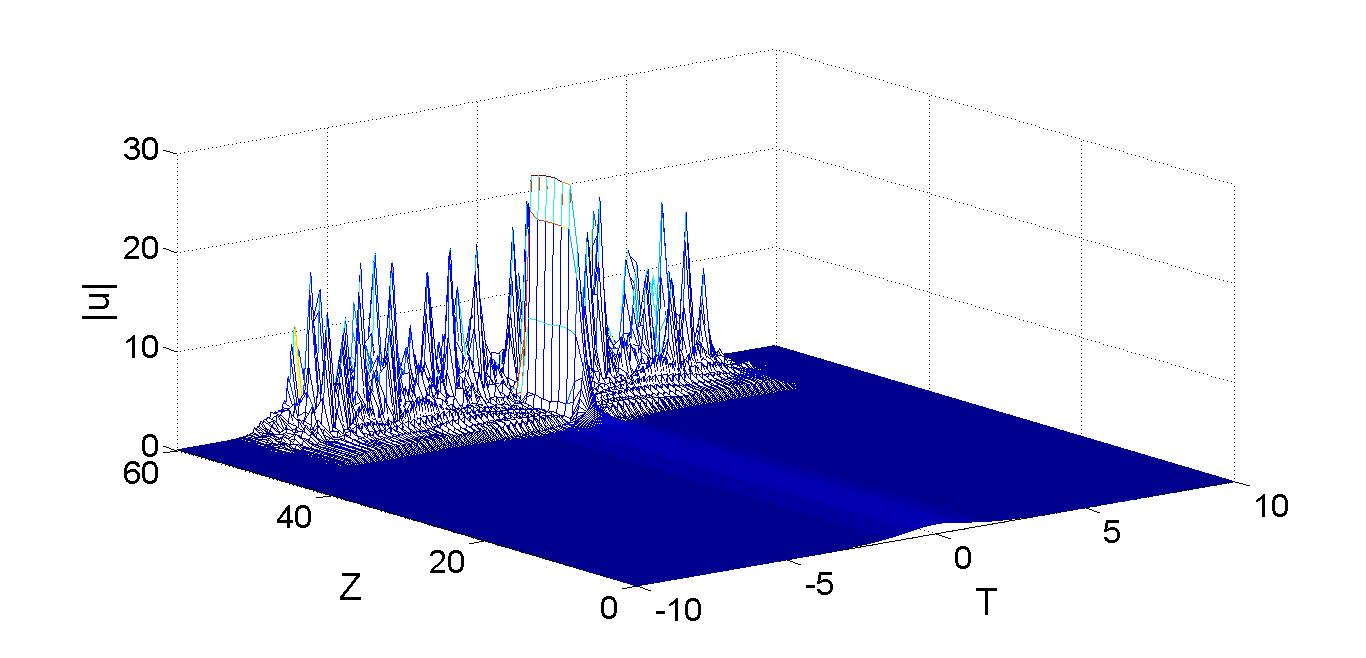}} %
\centerline{\includegraphics[width=7.5cm]{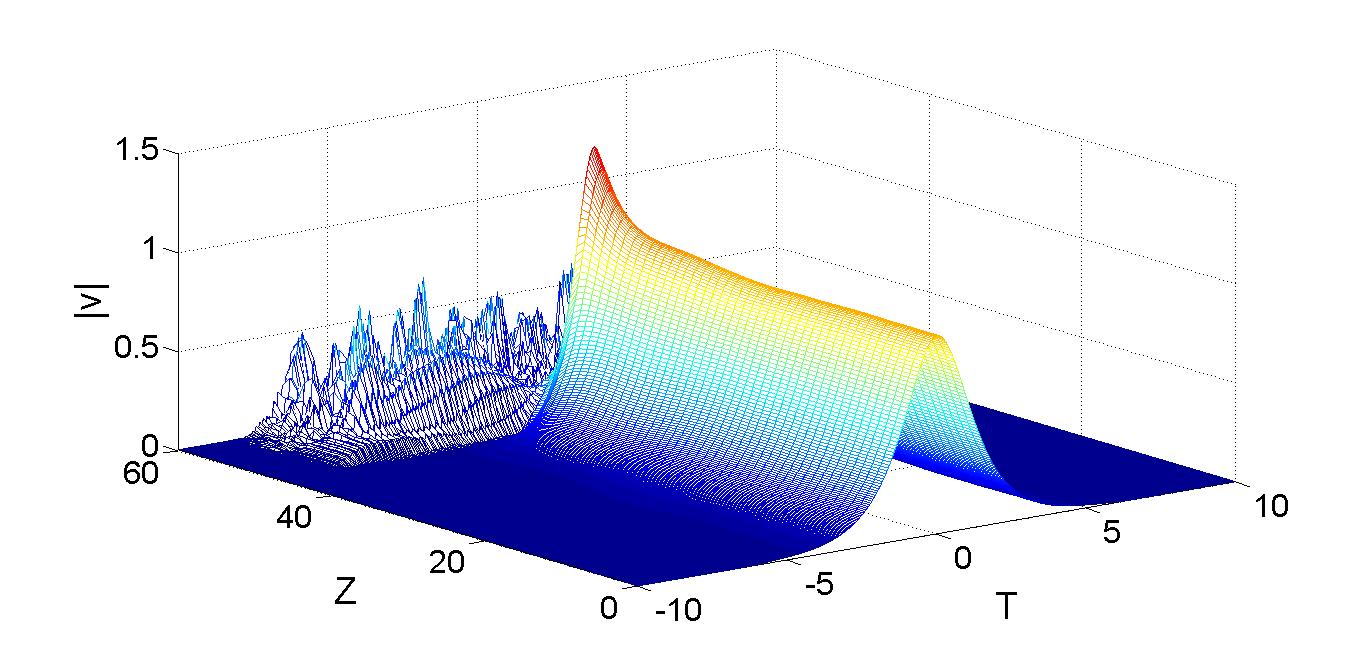}}
\caption{(Color online) Instability of a fundamental soliton in the
supersymmetric system (\protect\ref{uv1}) with $\protect\kappa =1$. The
solution is generated by the input corresponding to soliton (\protect\ref%
{sol}) with a small perturbation, $u_{0}(T)=-iv_{0}(T)=1.03\mathrm{sech}(T).$
Panels (a) and (b) display the evolution of the wave components in the
amplified and damped cores, respectively (note the difference in the scales
of the vertical axes between (a) and (b)).}
\label{fig1}
\end{figure}

The linearization of Eqs. (\ref{uv1}) around any exact solution of the form $%
v=iu$ leads to the following equations for small perturbations $\delta u$
and $\delta v$:
\begin{equation}
\mathcal{L}\left( \delta u+i\delta v\right) =0,~\mathcal{L}\left( \delta
u-i\delta v\right) =2i\kappa \left( \delta u+i\delta v\right) ,  \label{L}
\end{equation}%
where $\mathcal{L}\delta u\equiv \left[ i\partial _{Z}+(1/2)\partial
_{TT}+2|u|^{2}\right] \delta u+u^{2}\delta u^{\ast }$ corresponds to the
linearization of the NLS equation. If $u(Z,T)$ is stable as the solution to
the NLS equation (the fundamental soliton definitely is), the first equation
in system (\ref{L}) produces no instability, while a resonance is
anticipated in the second, inhomogeneous, equation, as $\delta u+i\delta v$
is an eigenmode (corresponding to of the same operator $\mathcal{L}$ which
figures on the right-hand side of equation. As follows from the commonly
known linear-resonance theory, perturbations generated by the latter
equation grow linearly in $Z$, rather than exponentially. Simulations of
Eqs. (\ref{uv1}) confirm that all the solitons are unstable, the character
of the instability development being consistent with the subexponential
growth, see Fig. \ref{fig1}.

As said above, we aim to develop the management method for stabilizing
solitons in the \textquotedblleft supersymmetric" system, which amounts to
periodically switching the gain and loss between the cores, with the
simultaneous reversal of the sign of the coupling coefficient. The
accordingly modified equations (\ref{uv1}) are

\begin{gather}
iu_{Z}+(1/2)u_{TT}+|u|^{2}u  \notag \\
+\kappa \mathrm{sgn}\left\{ \sin \left( 2\pi Z/L\right) \right\} \left(
-iu+v\right) =0,  \notag \\
iv_{Z}+(1/2)v_{TT}+|v|^{2}v  \notag \\
+\kappa \mathrm{sgn}\left\{ \sin \left( 2\pi Z/L\right) \right\} \left(
iv+u\right) =0,  \label{uv2}
\end{gather}%
where $L$ is the management period. We stress that the management does not
break the supersymmetry of the system, and, accordingly, wave forms with $%
v(Z,T)=iu(T,v)$, where $u$ is any solution to the usual NLS equation, remain
exact solutions of system (\ref{uv2}).

The stabilizing effect of the management can be explained by its action on
perturbations around the exact solutions. In particular, the replacement of
constant $\kappa $\ in the second equation in system (\ref{L}) by the
periodically flipping coefficient detunes the destabilizing resonance. The
detuning is small for a large management period, hence the stabilization
effect is expected to be weak in this case, which is indeed observed, see
Fig. \ref{fig3} below. Previously, effects of the periodic modulation of the
coupling strength on the stability of solitons were studied in the model of
the usual (conservative) dual-core systems \cite{Sydney,book}. Those effects
may be rather involved, but, generally speaking, they do not stabilize the
symmetric solitons against the symmetry-breaking bifurcation, as rapid
modulations effectively weaken the linear coupling between the cores, thus
making the symmetric solitons less robust against the spontaneous symmetry
breaking. Thus, the effects of the management mechanism acting on the
\textquotedblleft supersymmetric" solitons are different from the earlier
studied ones.

\begin{figure}[tbp]
\centerline{\includegraphics[width=8.2cm]{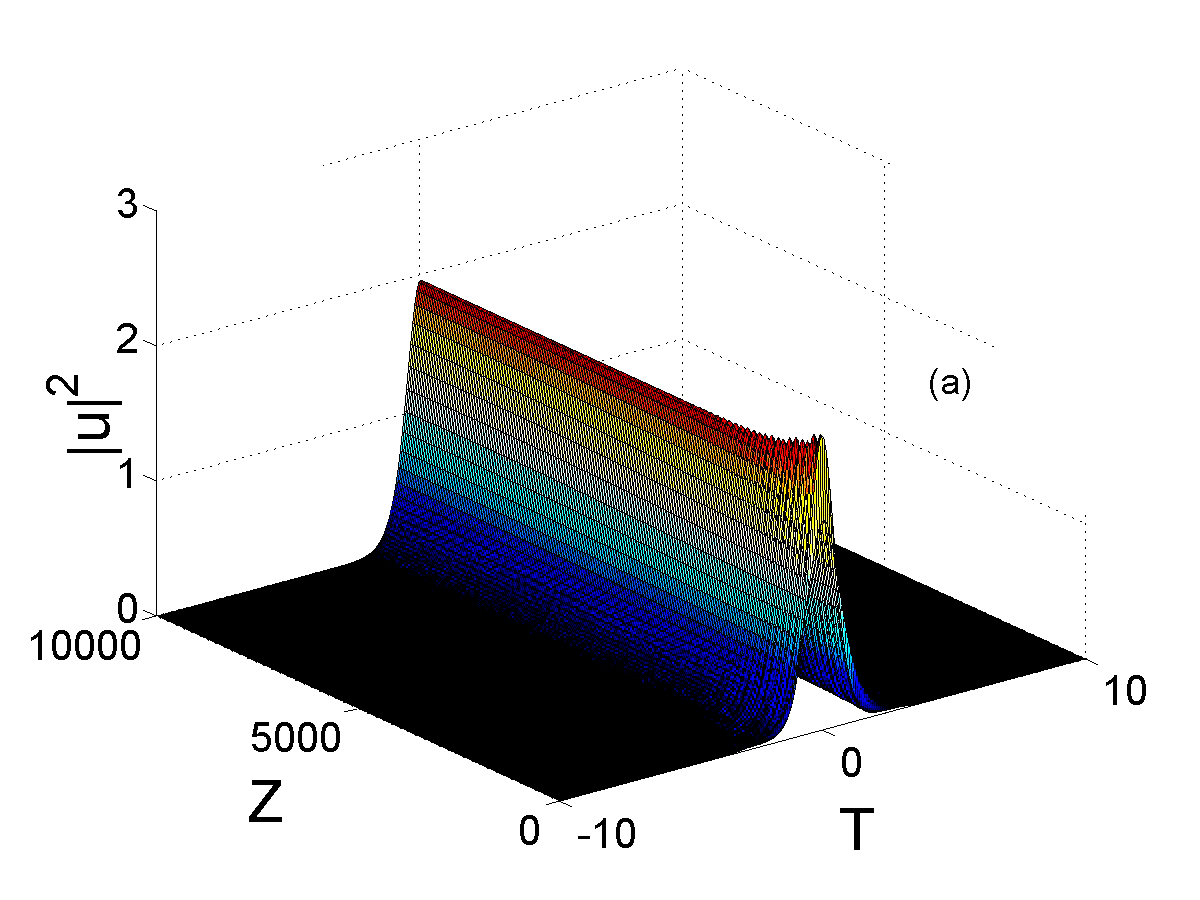}} %
\centerline{\includegraphics[width=8.2cm]{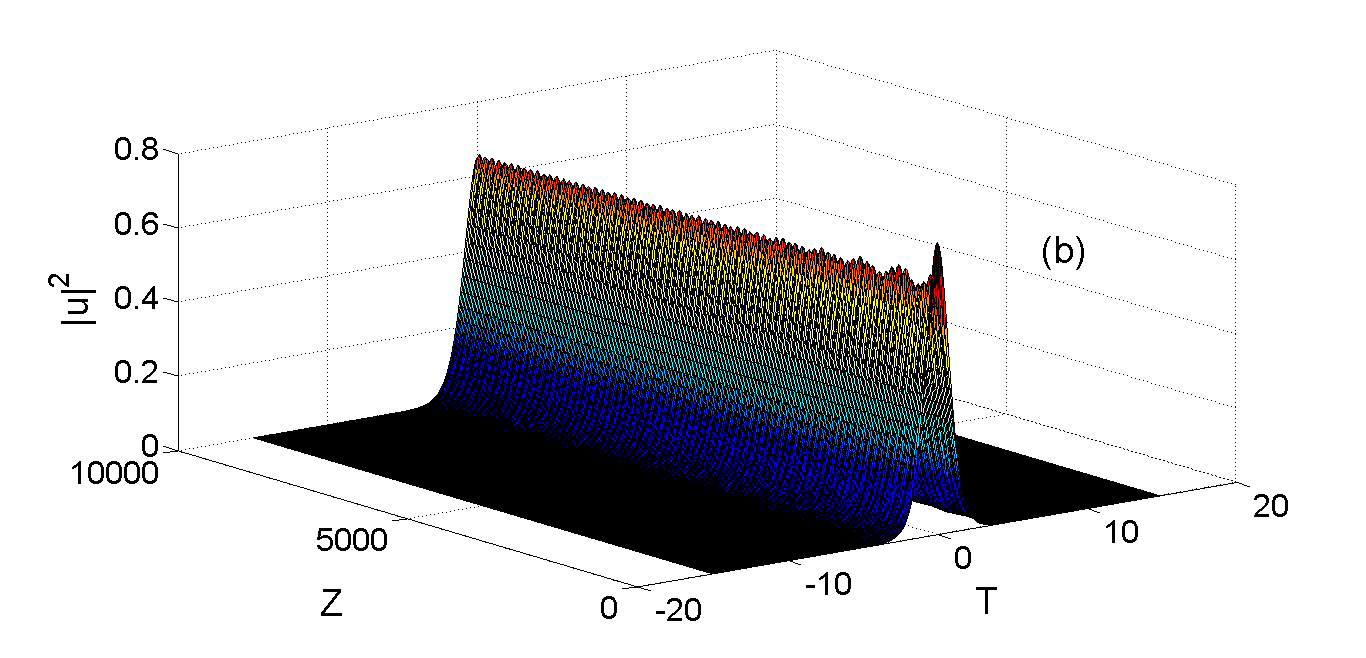}}
\caption{(Color online) The self-trapping of inputs (\protect\ref{in}) with $%
A=1.2$ (a) and $A=0.9$ (b) into stable solitons, demonstrated by simulations
of Eqs. (\protect\ref{uv2}) for $\protect\kappa =1$ and $L=2$. The evolution
of the $v$-component is similar to that displayed here for $u(Z,T)$.}
\label{fig2}
\end{figure}

\textit{The stabilization of solitons by means of the management}.
Simulations of Eqs. (\ref{uv2}) reveal that the periodic application of the
switching not only stabilizes the exact soliton solutions against small
perturbations, but actually turns the solitons into strong \textit{attractors%
} for inputs significantly different from the exact solutions. Typical
examples illustrating the rapid convergence of the inputs, taken as%
\begin{equation}
u_{0}(T)=A~\mathrm{sech}(T),v_{0}=iu_{0},  \label{in}
\end{equation}%
with amplitudes $A$ differing by $+20\%$ or $-10\%$ from the value $A_{0}=1$
in the respective exact soliton solution (\ref{sol}), into solitons are
displayed in Fig. \ref{fig2}. Note that the propagation distance shown here
in units of $Z$ actually coincides, within the order of magnitude, with that
measured in units of the soliton's dispersion length, therefore the
propagation distance displayed in the figure is extremely large, making the
full stabilization of the solitons obvious.

While the stabilization seems rather obvious in the case of the short-period
management (as in that case the frequent switch between the loss and gain
may nearly average them to zero), the model provides conspicuous
stabilization even for very large management periods. The results of
systematic simulations are summarized in Fig. \ref{fig3}, which displays
stability domains (areas between the curves) in the plane of the management
period $L$ and input amplitude $A$ (defined as per Eq. (\ref{in})), for
different values of the coupling parameter $\kappa $ in Eqs. (\ref{uv2}).
While it is natural that the stability area expands with the decrease of $%
\kappa $ (which corresponds to weaker perturbations), it is worthy to note
that the stability regions remain appreciable even for very strong coupling (%
$\kappa =10$). Note that (as was mentioned above) the stability areas may be
realized as attraction basins for the stable solitons which self-trap from
input pulses (\ref{in}).

\begin{figure}[tbp]
\centerline{%
\includegraphics[width=8.6cm]{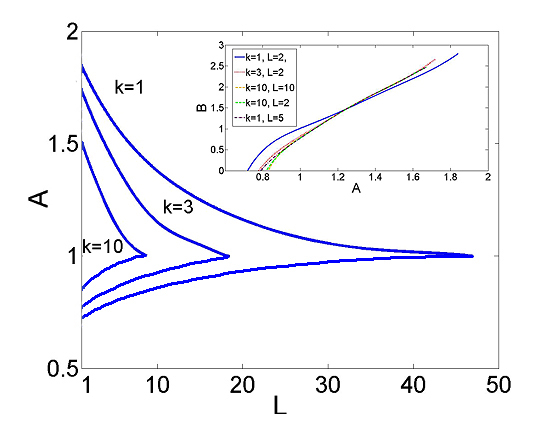}}
\caption{(Color online) The soliton stability regions in the $(L,A)$ plane
for different values of coupling parameter $\protect\kappa $ (see Eqs. (%
\protect\ref{uv2}) and (\protect\ref{in})). The inset shows amplitude $B$ of
the emerging stable soliton as a function of input amplitude $A$.}
\label{fig3}
\end{figure}

Amplitude $B$ of the self-trapped soliton, that has the form of $\left\vert
u(T)\right\vert =|v(T)|=B~\mathrm{sech}\left( BT\right) $, is a function of
the input's amplitude $A$ of initial condition (\ref{in}), as shown in the
inset of Fig. \ref{fig3}. It is worthy to note that, unlike the stability
area, dependence $B(A)$ is weakly affected by the variation of $L$ and $%
\kappa $.

\begin{figure}[tbp]
\centerline{\includegraphics[width=7.5cm]{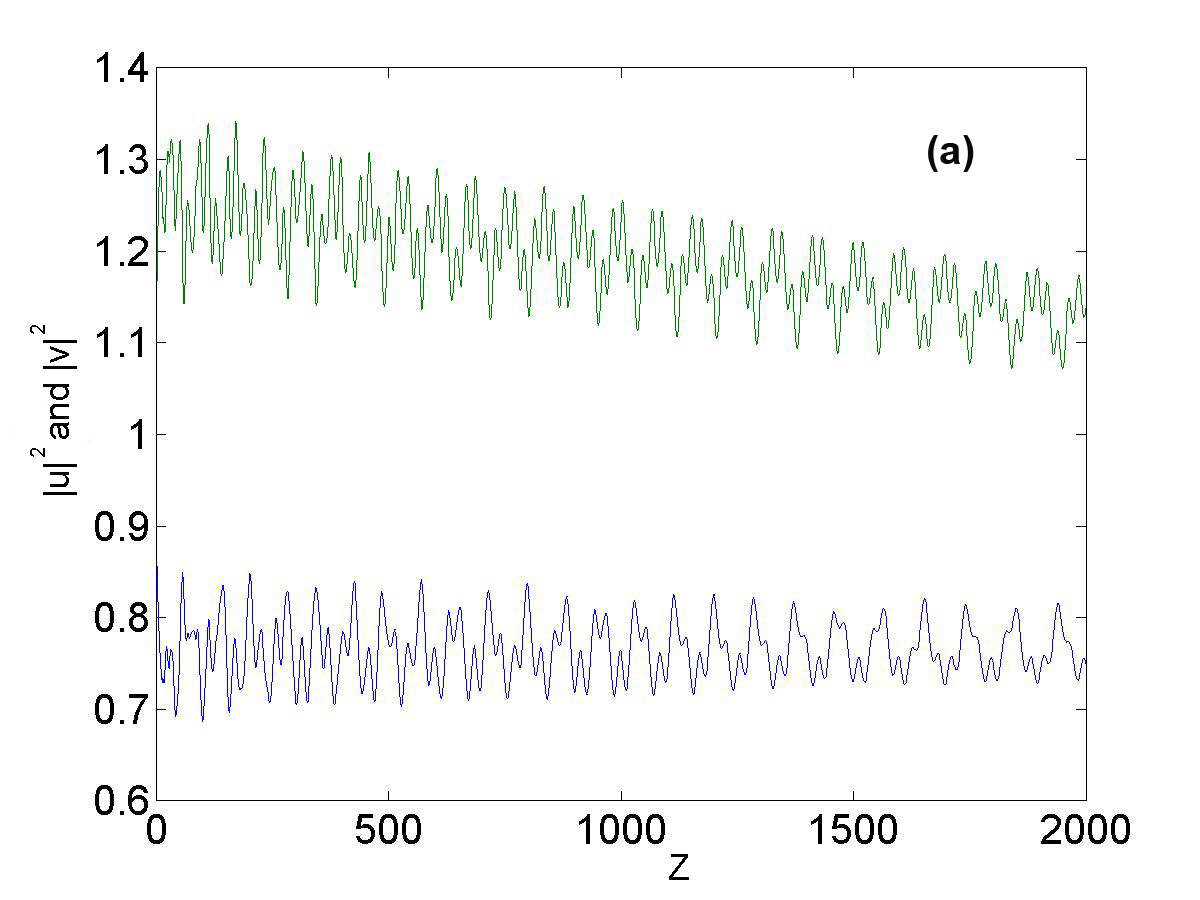}} %
\centerline{%
\includegraphics[width=7.5cm]{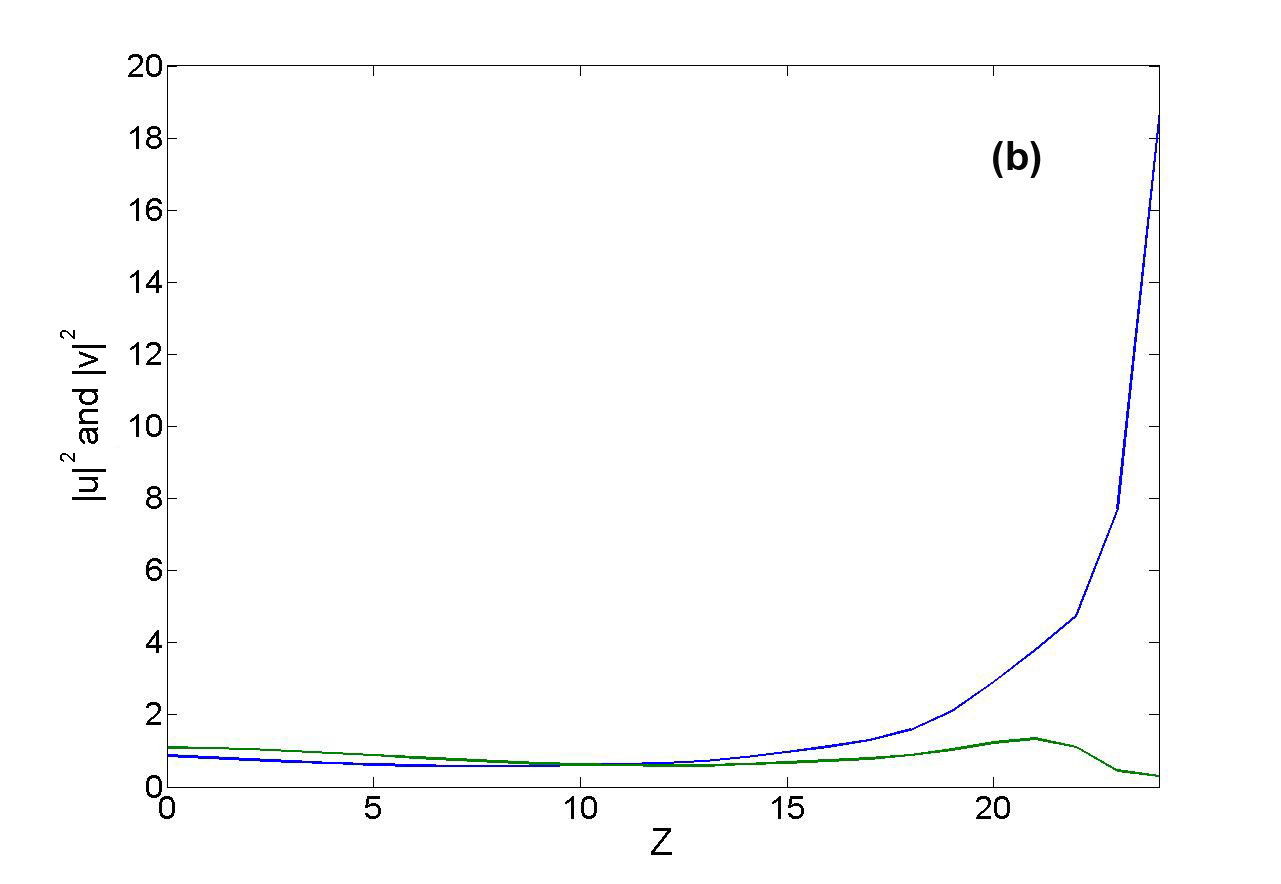}}
\caption{(Color online) The evolution of the peak powers of both field
components in the case of the initial conditions with asymmetric amplitudes,
$u_{0}=1.05\mathrm{sech}(T)$, $v_{0}=0.95i~\mathrm{sech}(T)$. The coupling
parameter is $\protect\kappa =0.25$. (a) The system under the action of the
management with period $L=2$. (b) The system without the management.}
\label{fig4}
\end{figure}

Obvious peculiarities of the exact supersymmetric solutions, with $v=iu$,
are the equality of the amplitudes of the two components, and the fixed
phase shift between them, $\pi /2$. The use of inputs which break either of
these relations impedes the self-trapping into exact solitons. In such a
case, the management gives rise to overall-stable two-component solitary
\textit{breathers}, which feature persistent irregular oscillations and
extremely slow evolution of the average peak power. A typical evolution
history produced by the input with the asymmetry in amplitudes is
illustrated by Fig. \ref{fig4}(a). For comparison, Fig. 4(b) shows that the
same input pulse quickly collapses in the system without management.

\begin{figure}[tbp]
\centerline{\includegraphics[width=8.2cm]{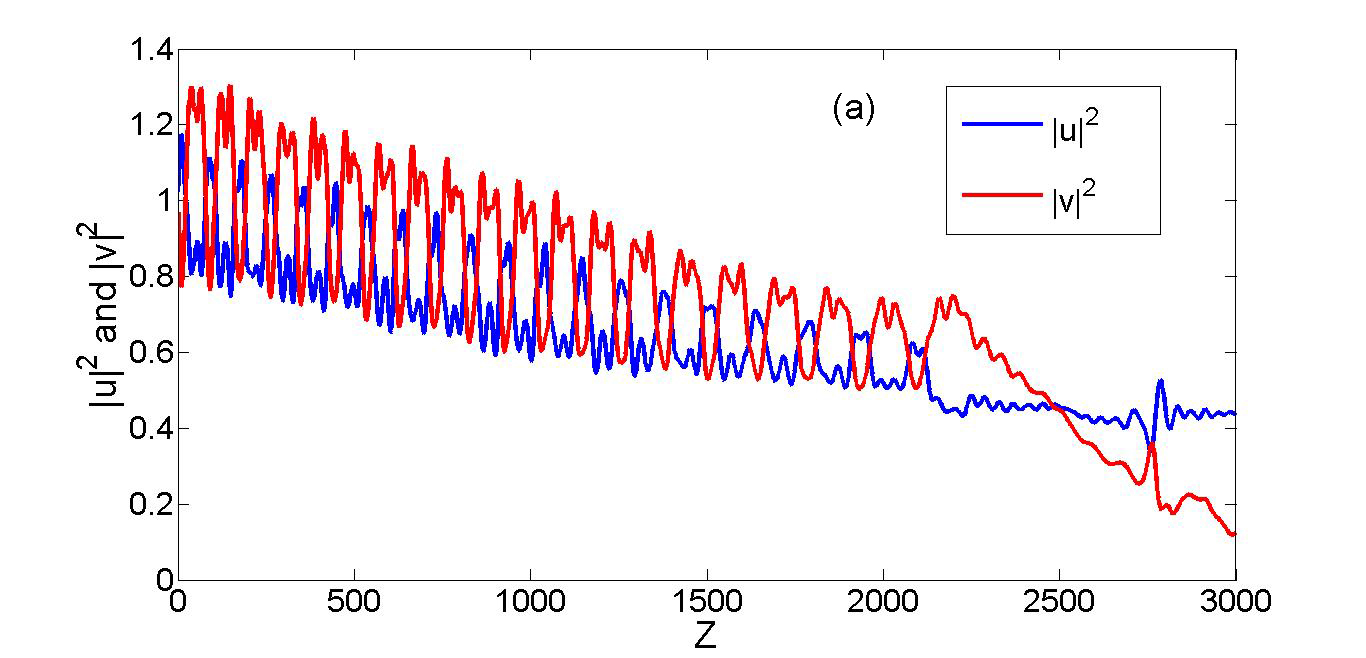}} %
\centerline{\includegraphics[width=8.2cm]{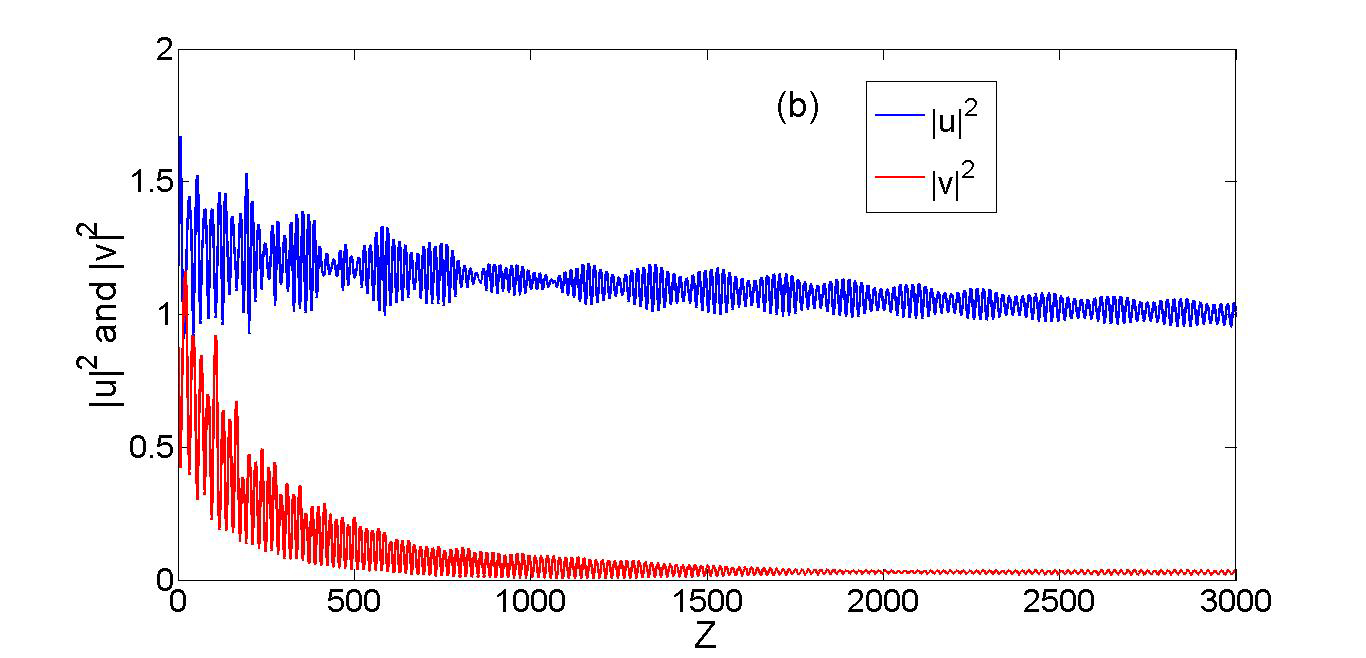}}
\caption{(Color online) The same as in Fig. \protect\ref{fig4}, but for $%
\protect\kappa =1$ and initial condition (\protect\ref{phase}) with phase
mismatch between the components: $\protect\phi =0.25\protect\pi $ in (a) and
$\protect\phi =0.5\protect\pi $ in (b).}
\label{fig5}
\end{figure}

The solitons in the supersymmetric system are also sensitive to a mismatch
in the phase shift between the two components. The simulations with input
\begin{equation}
u_{0}(T)=\mathrm{sech}(T),v_{0}=i\mathrm{\exp }(i\phi )u_{0}(T),
\label{phase}
\end{equation}%
where $\phi $ is the phase mismatch (cf. Eq. (\ref{sol})), generate
quasi-stable two-component breathers, which remain coupled over a finite but
long propagation distance, if $\phi $ is not too large. An example for $\phi
=0.25\pi $ is displayed in Fig. \ref{fig5}(a). On the other hand, the
increase of $\phi $ in the same case to $\phi =0.5\pi $ leads to quick
decoupling of the components and destruction of the two-component soliton,
as shown in Fig. \ref{fig5}(b).

\begin{figure}[tbp]
\centerline{\includegraphics[width=8.2cm]{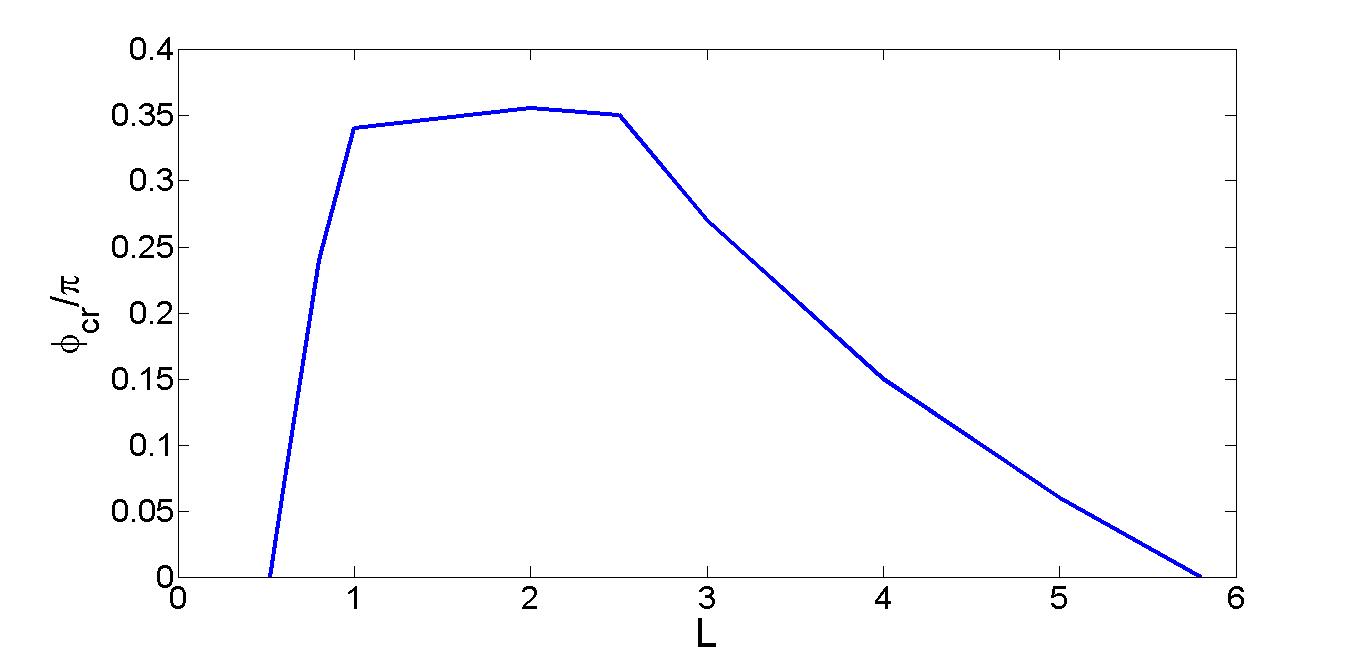}}
\caption{(Color online) The critical value of the initial phase mismatch, $%
\protect\phi _{\mathrm{cr}}$, for the input (\protect\ref{phase}), versus
the management period, $L$, for $\protect\kappa =1$ in Eq. (\protect\ref{uv2}%
).}
\label{fig6}
\end{figure}

Collecting results of the simulations, it is possible to identify a critical
value of the mismatch, $\phi _{\mathrm{cr}}$, which separates the
quasi-stable transmission of the two-component soliton and its quick
splitting. Fixing $\kappa \equiv 1$, $\phi _{\mathrm{cr}}$ is displayed as a
function of the management period, $L$, in Fig. \ref{fig6}. It is observed
that the phase mismatch produces the strong destabilization effect at small
and large values of $L$, while a flat optimum region is identified at $%
1\lessapprox L\lessapprox 2.5$. For $L<1$, detailed consideration of the
numerical results explains the destabilization as follows: small $L$
affects, first of all, the coupling between the components, practically
nullifying it due to the averaging over frequent flips of the sign of $%
\kappa $. Thus, the system splits into almost decoupled components, and the
one which starts its evolution under the action of the loss cannot recover
after the initial decay, despite the subsequent switch of the loss into the
gain (similar to what is seen in Fig. \ref{fig5}(b)). In the opposite case
of large $L$, the splitting of the components is simply explained by the
fact that one of them suffers strong attenuation during the first period of
the management.

\textit{Conclusion}. We have introduced a system based on the dual-core
coupler with the intrinsic Kerr nonlinearity, which features the
supersymmetry, as an extended version of the $\mathcal{PT}$ symmetry, with
the gain and loss in the two cores equal to the strength of the inter-core
coupling. The supersymmetry allows one to find a broad class of exact
solutions, including solitons, for which the stability is a crucial issue.
The exact solutions are vulnerable to the specific subexponential
instability, which destroys all the solitons. We have demonstrated that the
management, defined as periodic synchronous switch of the signs of the gain,
loss, and linear coupling readily stabilizes the solitons, keeping the
system's supersymmetry intact. Actually, the management makes solitons
attractors, for which we have identified the attraction basins. The
amplitude asymmetry and phase mismatch between the two components in the
input pulse turns the stationary solitons into quasi-stable breathers. We
have identified the stability area for the breather with respect to the
phase mismatch and management period.

As an extension of this study, it may be relevant to consider interactions
between the stabilized solitons. A challenging problem is to analyze
two-dimensional double-core supersymmetric systems.

\end{document}